\documentclass[structabstract]{aa}

\usepackage{epsfig}
\usepackage{graphicx}
\usepackage{txfonts}

\begin{document}

\title{Far-infrared spectra of hydrous silicates at low 
temperatures\thanks{Based on observations with ISO, an ESA project 
with instruments funded by ESA Member States (especially the PI 
countries: France, Germany, the Netherlands, and the United Kingdom) 
and with the participation of ISAS and NASA.}}
\subtitle{Providing laboratory data for Herschel and ALMA}
\author{H.Mutschke\inst{1}, S. Zeidler\inst{1}
\and
Th. Posch\inst{2}, F. Kerschbaum\inst{2}, A. Baier\inst{2}
\and
Th. Henning\inst{3}}

\institute{Astrophysikalisches Institut,
Schillerg\"asschen 2-3, D-07745 Jena, Germany\\
\email{mutschke@astro.uni-jena.de, sz@astro.uni-jena.de}
\and
Institut f\"ur Astronomie, T\"urkenschanzstra{\ss}e 17, 
A-1180 Wien, Austria\\
\email{posch@astro.univie.ac.at}
\and
Max-Planck-Institut f\"ur Astronomie (MPIA), K\"onigstuhl 17, 
D-69117 Heidelberg, Germany\\
\email{henning@mpia.de}}

%-----------------------------------------------------------------------------

\date{Received date; accepted date}

\abstract
 %context heading (optional)
{Hydrous silicates occur in various cosmic environments, and 
are among the minerals with the most pronounced bands in the 
far infrared (FIR) spectral region. 
Given that {\em Herschel}\/ and ALMA will open up new possibilities
for astronomical FIR and sub-mm spectroscopy, data characterizing the 
dielectric properties of these materials at long wavelengths 
are desirable.}
 % aims heading (mandatory)
{We aimed at examining the FIR spectra of talc, picrolite, 
montmorillonite, and chamosite, which belong to four different groups 
of phyllosilicates. We tabulated positions and band widths of the FIR 
bands of these minerals depending on the dust temperature. }
 % methods heading (mandatory)
{By means of powder transmission spectroscopy, spectra of the examined
materials were measured in the wavelength range 25--500\,$\mu$m 
at temperatures of 300, 200, 100, and 10\,K.}
 % results heading (mandatory)
{Room-temperature measurements yield the following results. 
For talc, a previously unknown band, centered at 98.5\,$\mu$m,
was found, in addition to bands at 56.5 and 59.5\,$\mu$m.
For montmorillonite, several bands at wavelengths $<$110\,$\mu$m 
were detected, including a band at 105\,$\mu$m with an FWHM of about 
10\,$\mu$m. Picrolite shows a sharp 77\,$\mu$m FIR band. 
Chamosite is characterized by bands in the 72--92\,$\mu$m range, 
and a prominent band at 277\,$\mu$m. At decreasing temperature, 
most of the bands shift to shorter wavelengths.}
% conclusions
{Examining a potential counterpart of the 105$\mu$m band in the 
spectra of HD 142527 and HD 100546, we find that the broad band in the 
spectra of these young stars -- extending from 85 to 125\,$\mu$m -- 
cannot be due to montmorillonite or any of the hydrous silicates we 
studied, since these materials have sharper bands in the FIR 
wavelength range than previously assumed, especially at low 
temperatures.}

\keywords{stars: circumstellar matter --- stars: planetary systems: protoplanetary disks --- infrared: stars --- methods: laboratory}

%------------------------------------------------------------------------------

\maketitle

\section{Introduction \label{sec:intro}}

Phyllosilicates, which are partly called hydrous silicates due
to the inclusion of OH groups in their lattice structures, are an 
important constituent of primitive meteorites, such as the carbonaceous 
chondrites of the CM and CI classes (e.g.\,Nagy 1975), and of a 
corresponding class of interplanetary dust particles, the chondritic 
smooth IDPs (e.g.\ Smith et al.\ 2005, Sandford \& Walker 1985). 
They are probably formed by aqueous alteration of pyroxene and 
olivine minerals in the parent bodies of these meteorites, also an 
important weathering mechanism for terrestrial rocks. 
For example, terrestrial enstatite (MgSiO$_3$) is often converted 
to talc Mg$_3$[Si$_4$O$_{10}|$(OH)$_2$] by weathering (see, 
e.g.\ J\"ager et al.\ 1998). Also, forsterite (MgSiO$_4$) and 
tremolite Ca$_2$Mg$_5$Si$_8$O$_{22}$(OH)$_2$ can easily be 
transformed into talc. Terrestrial talc is often associated 
with serpentine, tremolite, forsterite, and almost always with 
carbonates. It is transformed to enstatite at temperatures
above $\sim$1000K (Kr\"onert et al.\ 1964).

In a number of cases, infrared spectroscopy has provided hints 
to the presence of phyllosilicates in astrophysical objects. 
Zaikowski, Knacke, \& Porco (1975), Dorschner et al.\ (1978), and 
Knacke \& Kr\"atschmer (1980) were among the first to measure 
phyllosilicate mid-infrared spectra, because the phyllosilicate 
bands roughly corresponded to the absorption properties of interstellar 
silicate dust. They found, however, that all the phyllosilicates had 
much too narrow bands as to account for the observational data. 
Later it became clear that amorphous silicates instead of crystalline 
phyllosilicates can account for most of the $\sim$10\,$\mu$m spectra 
of oxygen-rich cosmic dust. 

Nevertheless, after the ``crystalline silicate revolution'' with the 
increasing quality and wavelength coverage of the observational data, 
new indications for phyllosilicate bands have been found in the infrared 
spectra of cometary dust (e.g.\ Lisse et al.\ 2007), young stellar 
objects such as HD 142527 (Malfait et al.\ \cite{M99}), and planetary 
nebulae (Hofmeister \& Bowey 2006). Even if these identifications are 
based on either extremely complex fits or on the assignment of a small 
number of individual IR bands and may therefore be a matter of debate, 
the inclusion of phyllosilicates into dust models may be a natural step 
for various circumstellar environments. 

For protostellar or protoplanetary systems, the above-mentioned 
capacity of phyllosilicates to integrate OH-groups, or even H$_2$O molecules, 
into their structure makes them a unique potential water reservoir. 
In our own solar system, hydrosilicates may have even been the main source 
for liquid water on Earth. According to present knowledge, fragments of 
asteroids contained even enough crystal water to create the oceans 
when colliding with the young planet. 

Phyllosilicates show infrared bands over a very wide wavelength range.
This is due to their internal structure which enables different kinds
of vibrations from O--H stretching modes in the near infrared to lattice modes
in the far infrared (FIR). Koike \& Shibai (1990), e.g., have published optical 
constants and extinction efficiencies for chlorite (among other hydrous silicates),
including a peak even at 268\,$\mu$m. Most of the characteristic phyllosilicate
bands that distinguish these minerals from structurally less complex silicates
are, however, located in the 25--100\,$\mu$m wavelength domain.

Upcoming astronomical facilities like {\em Herschel}\/ will probe
the FIR domain in unpreceded detail with
a spectral resolution ideally suited for mineralogical studies.
The spectrometers in {\em Herschel}\/-PACS and -Spire will cover
ranges from 55--210\,$\mu$m and 200--670\,$\mu$m, respectively.
Posch at al.\ (2005) gave an overview of some FIR solid-state features
and pointed out the need for laboratory work. 
ALMA will extend these ranges to even longer wavelengths.
Thus it is highly desirable to enlarge the laboratory data sets of 
cosmic dust species to the FIR.

The spectroscopic work presented in this paper aims 
at an extension and specification of previous FIR data including 
systematic low temperature measurements for a sequence of cryogenic 
temperatures. Before discussing our data, we briefly summarize the 
results of previous papers -- mostly from the astrophysical 
literature -- on FIR data for the same group of minerals.

Koike, Hasegawa, \& Hattori (1982) measured infrared absorption spectra 
of chlorite, montmorillonite, and serpentine at wavelengths up to 115\,$\mu$m 
including spectra measured at a temperature of 2\,K. Furthermore, they 
investigated the thermo-metamorphism of these minerals by means of 
mid-infrared spectroscopy. Koike \& Shibai (1990) extended these 
data to 400\,$\mu$m for room temperature and derived optical constants 
of the three minerals from the transmission spectra. 
Thus, pre-existing sets of optical constants, published by Toon et al.\ (1977) 
and by Mooney \& Knacke (1985), could be extended far beyond 50\,$\mu$m, 
to which wavelength the latter data were limited. 

Based on these data, Malfait et al.\ (\cite{M99}) assigned a very broad 
emission shoulder in the ISO-LWS spectrum of the Herbig Ae/Be star 
HD 142527 to cold montmorillonite dust. Apart from montmorillonite, 
crystalline water ice, FeO, and amorphous silicates are assumed to be 
present in the dusty disk surrounding HD 142527. Whether or not the 
assignment of the broad $\sim$105\,$\mu$m shoulder highlighted by 
Malfait et al.\ (\cite{M99}) to phyllosilicate dust is also consistent 
with our laboratory data will be discussed in Sect.\ \ref{sec:astro} 
below. Analyzing UKIRT spectra of circumstellar shells and star forming regions,
Bowey \& Adamson (2002) conclude the presence of up to 10 percent talc
and clays in these dusty environments. However, Bowey \& Adamson 
referred exclusively to the fine structure of the 10\,$\mu$m
silicate band in their study. 

Hofmeister \& Bowey (2006) present detailed results of quantitative
IR spectroscopy of brucite (Mg(OH)$_2$), chain silicates, and phyllosilicates.
Among the last, they examined chrysotile, lizardite, talc, saponite, and
montmorillonite. Their measurements cover a large wavelength range (from 
2.5 to beyond 100\,$\mu$m) and were performed at room temperature.
An extensive discussion of the assignment of the phyllosilicate bands
to individual stretching, bending, and lattice modes is included in their
paper, such that we can be very brief on this point.

Zhang et al.\ (2006) studied the behavior of talc bands upon heating 
to 1373\,K in steps of 100--200\,K from a mineralogical point of view.
Their spectra nicely illustrate the transformation of talc into
amorphous SiO$_2$, orthoenstatite (MgSiO$_3$, space group Pbca), 
and clinoenstatite (MgSiO$_3$, space group P2$_1$/c) at temperatures 
close to 1273\,K, thus adding to the similar study by Koike et al.\ (1982) 
for serpentine, montmorillonite, and chlorite. 

This cursory literature overview may suffice to illustrate both the
relevance of phyllosilicates to cosmic dust research and the need 
for detailed spectroscopic studies in the FIR range, especially at the 
low temperatures at which these minerals presumably exist in space. 

The present paper is structured as follows: In Sect.\ \ref{sec:meth}, 
we mainly describe the experimental methods we used (including the 
cryogenic measurements). 
The subsequent Sect.\ \ref{sec:res} contains the FIR spectra
of talc, picrolite, chamosite, and montmorillonite. Finally, in
Sect.\ \ref{sec:astro}, we discuss the implications of our study for
astromineralogical investigations.

%------------------------------------------------------------------------------

\section{Experimental methods \label{sec:meth}}

\begin{table*}
\caption{Overview of our sample minerals, their origins and chemical
compositions. 
\label{t:chem}}
\begin{center}
\begin{tabular}{llll}
\hline
\centering
mineral name & ideal sum formula & actual sample composition & sample origin\\
\hline
picrolite       & Mg$_6$[Si$_4$O$_{10}|$(OH)$_{8}$]  
                  & Mg$_{5.84}$Fe$_{0.17}$[Si$_4$O$_{10}|$(OH)$_{8}$] 
                  & Black Lake Mine, Canada \\
talc            & Mg$_3$[Si$_4$O$_{10}|$(OH)$_{2}$]    
                  & Mg$_{3.33}$Fe$_{0.1}$[Si$_4$O$_{10}|$(OH)$_{2}$]
                  & Murcia, Spain \\
chamosite       & (Fe,Al)$_6$[(Si$_3$Al)O$_{10}|$(OH)$_{8}$] 
                  & Fe$_{3.55}$Al$_{1.88}$[Si$_3$AlO$_{10}|$(OH)$_{8}$] 
                  & St.Brigitte, France \\
montmorillonite & (Mg,Al)$_2$[Si$_4$O$_{10}|$(OH)$_{2}$] & Al$_{1.5}$Mg$_{0.25}$Fe$_{0.18}$[Si$_4$O$_{10}|$(OH)$_{2}$] & Wyoming (USA) \\
& (Na,K,Ca)$_x$\,.\,nH$_2$O  &  (Na,K)$_{0.06}$\,.\,1.2H$_2$O & \\
\hline
\end{tabular}
\end{center}
\end{table*}

The basic building blocks of phyllosilicates are continuous two-dimensional
sheets with the sum formula Si$_4$O$_{10}$, with SiO$_4$ tetrahedra arranged 
in a hexagonal network. The oxygen at the `top' or `bottom' of the individual 
tetrahedra (i.e., the one directed perpendicular to the sheet) forms part of an 
adjacent octahedral sheet where the octahedra are linked to each other via
common edges. The octahedral sheets typically consist of hydroxides, such as
brucite Mg(OH)$_2$ or gibbsite Al(OH)$_3$. 

In terms of idealized stereometric arrangements, two- and three-layer-structures 
can be distinguished. In the former, the octahedral sheets simply alternate with 
the tetrahedral ones. In the latter, the cations of the hydroxide layers 
(e.g.\ Mg$^{++}$) are linked both up- and downwards to the SiO$_4$-based sheets; 
the tetrahedral sheets that are `sandwiching' a given octahedral sheet are thus 
oriented in opposite directions. 

The minerals studied in this paper represent different subtypes of the 
basic phyllosilicate structure described above. {\em Picrolite}\/, a member of 
the serpentine group, represents the two-layer (or 1:1 layer) type. 
{\em Talc}\/ is a three-layer (or 2:1 layer) phyllosilicate.
{\em Montmorillonite}\/, too, has a three-layered structure, 
but it has the particularity that H$_2$O (as well as other polar molecules,
among them organic ones) can enter between the unit layers, thereby
leading to a strong volume increase of the mineral.
{\em Chamosite}\/, of the chlorite group, has a rather complex
structure. It is composed of talc-type- and hydroxide-layers;
sometimes, the chlorite structure is therefore referred to as a 
`2:2' layer structure (e.g., Nagy 1975). Both in the talc and
in the brucite sublayers, Mg is replaced by Fe and Al cations in 
chamosite. The general sum formulae of the four different minerals
that we studied are given in Table \ref{t:chem}, together with
actually measured chemical compositions.

Natural minerals usually contain impurities. Therefore, we carefully 
selected parts of the samples without inclusions and checked their 
crystal structure by X-ray diffraction. The mid-infrared spectra, 
as well, were examined for signatures of well-known impurities, such as 
quartz or carbonates, but with negative results (see also Fig.\,\ref{cha_mir}). 

The ratios of the metal ions and silicon atoms indicated in Table \ref{t:chem} 
were determined by energy-dispersive X-ray (EDX) analysis. 
Oxygen and OH are not accessible by this method, so only the theoretical 
numbers are given. The apparent nonstoichiometry in the formulae is due to 
the measurement uncertainty or may be partly due to replacement of silicon 
by metal ions. The water content of montmorillonite was estimated 
by measuring the mass loss upon annealing for 30\,min at 1050\,$^{\circ}$C. 
This mass loss was 9.6\%$\pm$0.5\%. Since we know from thermogravimetric 
measurements on related samples (``bentonite'' with montmorillonite the 
major constituent) that around 4\% of the mass are released during the 
structural changes occurring around 700\,$^{\circ}$C, we can estimate 
that the inter-layer water released typically below 150\,$^{\circ}$C 
amounts to a difference of about 5.5\% of the total mass. 
This corresponds to about 1.2 H$_2$O molecules per structural unit.

%.............................................................................

\subsection{Sample preparation}

The spectroscopic measurements were performed on fine-grained particulates 
produced by grinding the minerals in a silicon nitride ball mill or in 
an agate mortar (when the amount of sample was much less than 0.5\,g as 
in the case of montmorillonite), followed by sedimentation in acetone. 
The powder fraction not sedimented down after a certain time, as calculated 
(according to Stokes law) from the desired upper grain size limit of d=2$\,\mu$m, 
was dried at 80\,$^\circ$C and embedded into polyethylene (PE) pellets. 
Given that the crystal structure facilitates the production of oblate particle 
shapes by grinding, the sedimentation times may actually be longer than 
calculated for spherical grains. This may have led to the presence of grains 
up to diameters of 5$\,\mu$m in the samples. 

Each mineral sample was mixed with 200\,mg of PE in an agate mortar at mass 
ratios of 1:100 for overview spectra. For more sensitive measurements at 
longer wavelengths, additional pellets with a mass ratio of 1:10 were 
prepared. After homogenizing the mixtures for 20--30\,min under wet conditions 
with spectral clean ethanol, the mixture was dried in a compartment drier for 
30--40\,min at about 80\,$^\circ$C to evaporate the ethanol solvent. 
The dry powder was pressed to a pellet for 5\,min under 10\,t load. 
The pellet diameter was 13\,mm, resulting in material column densities of 
$\sigma$ = 1.49\,mg/cm$^2$ for samples embedded at a ratio of 1:100 and 
13.7\,mg/cm$^2$ for samples with a ratio of 1:10.

\subsection{Measurement procedure}

For recording the FIR transmission spectra, we used a Bruker 113v Fourier-transform 
infrared spectrometer equipped with a continuous flow helium cryostat 
(Cryo Vac KONTI Spectro B). In this cryostat, samples were cooled by helium 
gas at a pressure of 10\,mbar, providing thermal contact with the liquid-helium 
cooled sample chamber walls. The temperatures of sample and wall were controlled 
by two silicon diode temperature detectors. Heating of the sample mount allowed 
for stabilization of the sample temperature to less than 1\,K variation during 
the 20--30\,min measuring time at each temperature. The measurements were 
performed at 300\,K, 200\,K, 100\,K, and 10\,K. 

The sample mount of the cryostat provides two sample positions, which can be 
alternately placed into the transmission beam. One of these sample positions, 
which was either kept empty or filled with a blank PE pellet, was used for 
reference measurements that were taken at each temperature. In the former 
case (empty reference) separate measurements of the blank PE pellet were 
performed to correct for losses due to PE reflection, absorption, 
and scattering. This approach allowed for an easier correction of fringes by 
avoiding superposition of sample and reference pellet fringes. 

For our measurements we used a globar as radiation source and mylar beam 
splitters of 3.5\,$\mu${m} and 12\,$\mu{m}$ thickness for the wavenumber 
ranges 1/$\lambda$ = 660--150\,cm$^{-1}$ and 220--50\,cm$^{-1}$, respectively. 
A mercury lamp and a 23\,$\mu{m}$ thick mylar beamsplitter have been used for 
the wavenumber range 120--30\,cm$^{-1}$. The transmitted IR beam was recorded 
by a DTGS-Detector (deuterated triglycine sulfate) with a PE window. 
The spectral resolution was chosen to amount to 2\,cm$^{-1}$ in the higher 
wavenumber ranges and 0.5\,cm$^{-1}$ in the lower wavenumber range.

%------------------------------------------------------------------------------

\section{Results \label{sec:res}}

\subsection{Spectra at lower column density and experimental limits}

\begin{figure*}
\centering
\includegraphics[width=9cm]{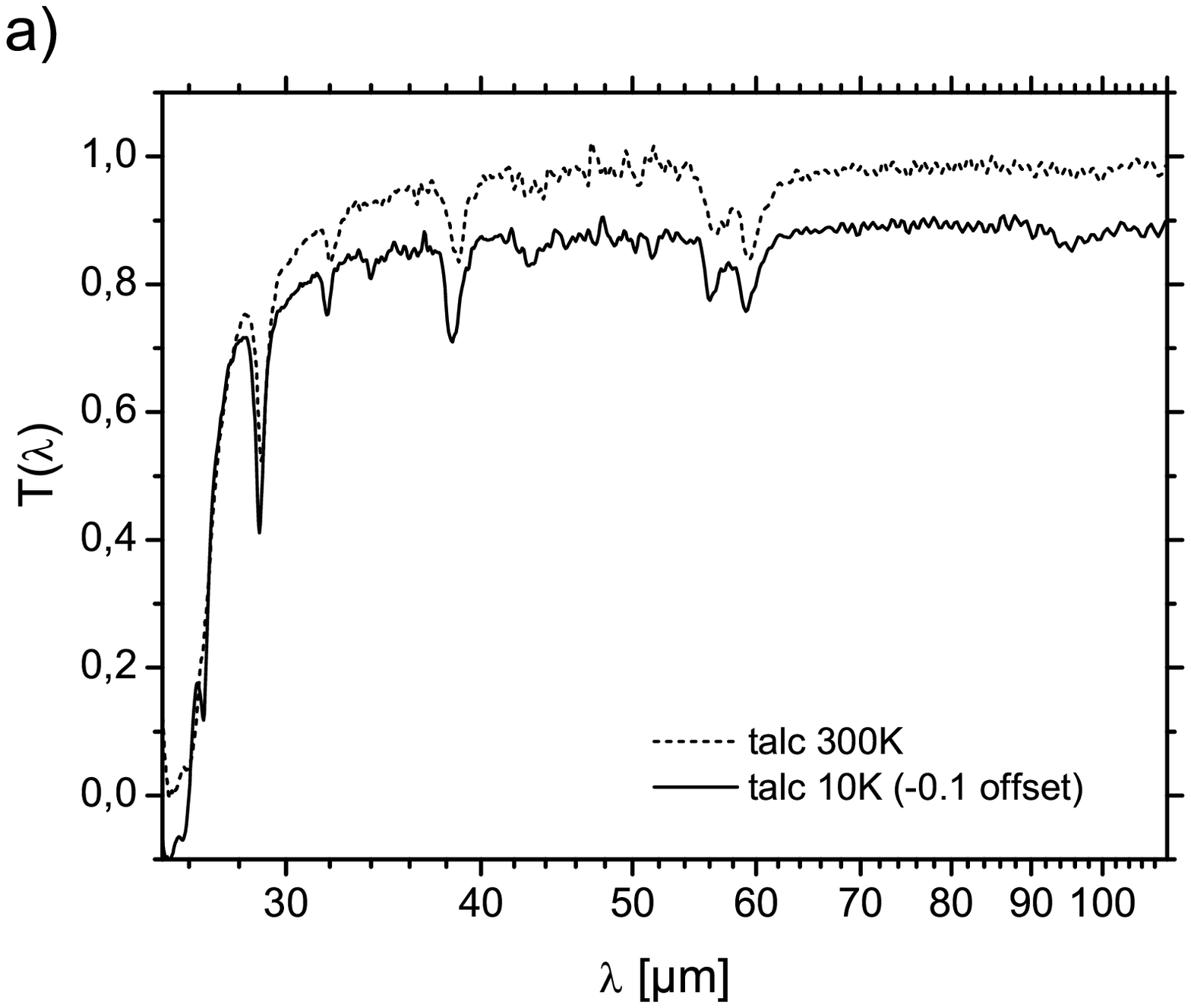}
\includegraphics[width=9cm]{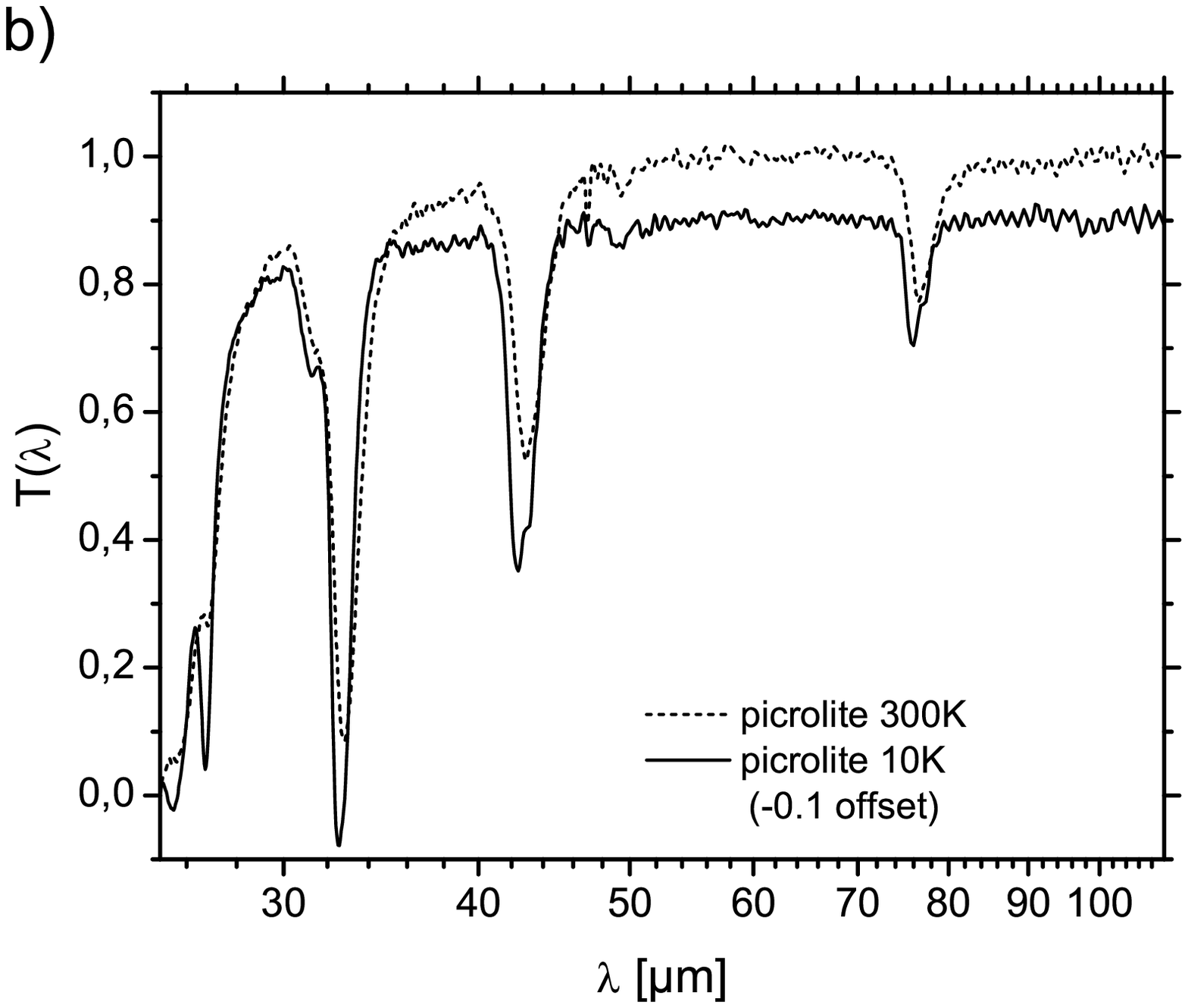}
\includegraphics[width=9cm]{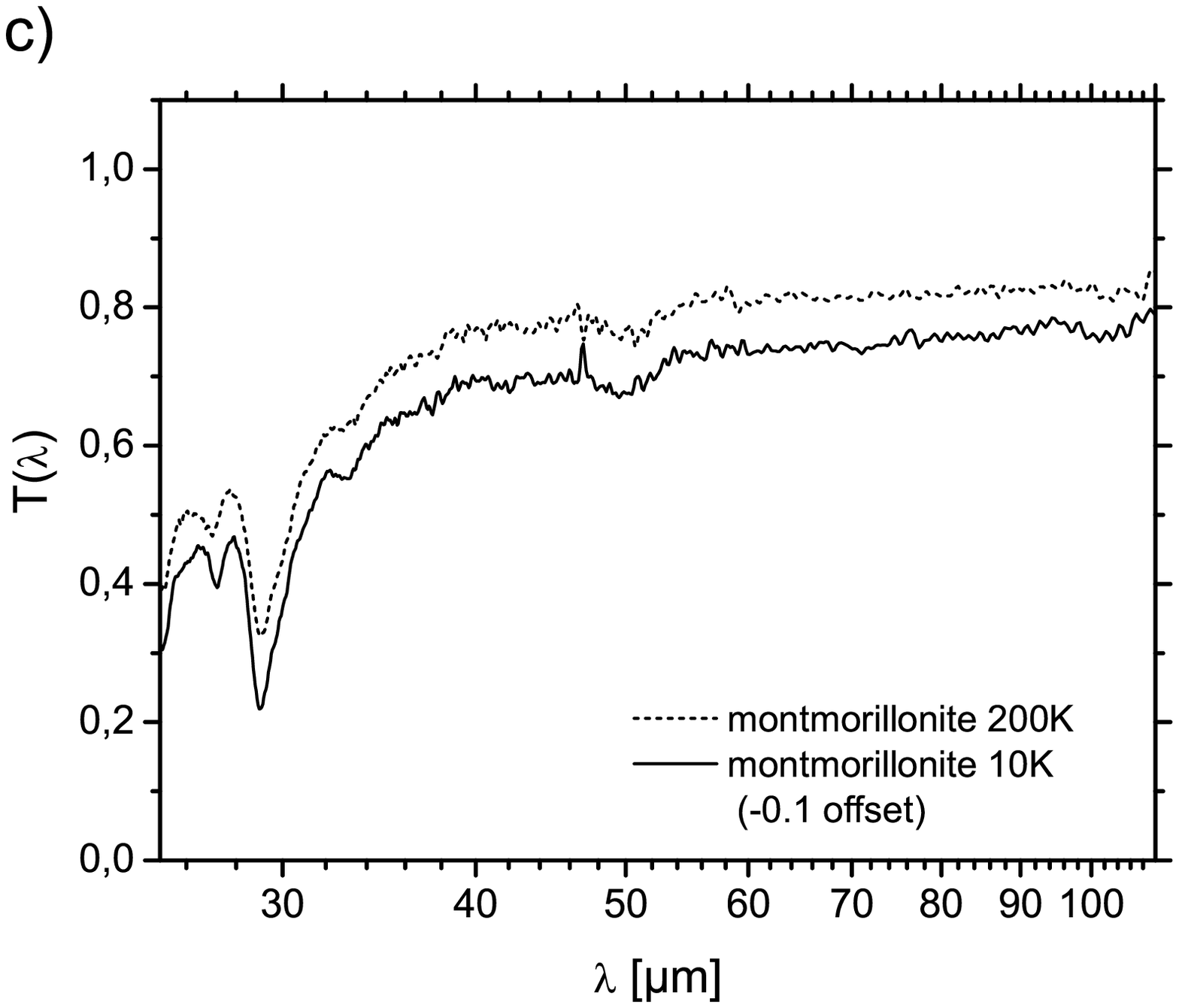}
\includegraphics[width=9cm]{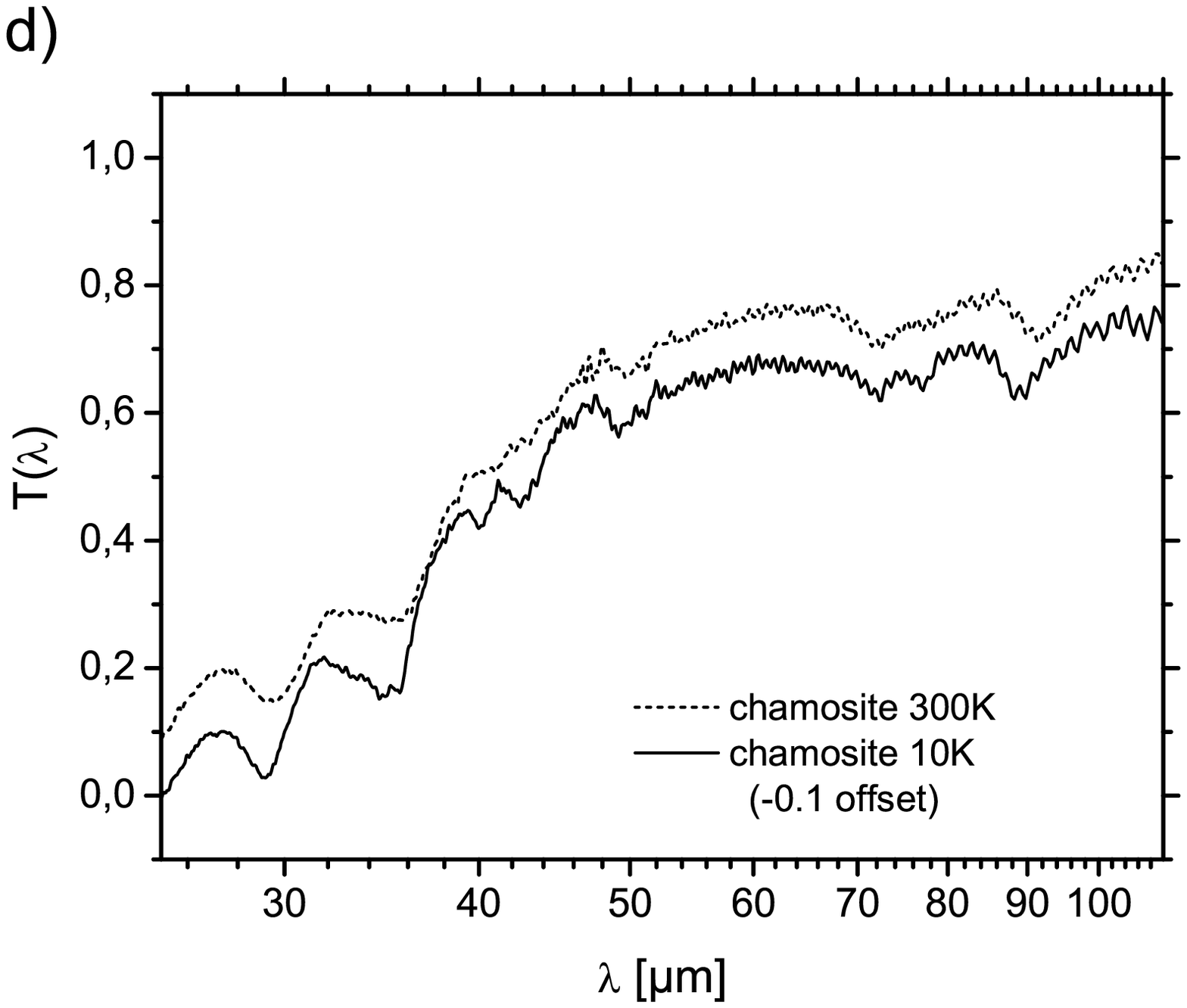}
\caption{Overview of the transmission spectra of (a) talc, (b) picrolite,
(c) montmorillonite, and (d) chamosite in the 25--110\,$\mu$m region. 
The dotted lines refer to measurements at room temperature, while the solid 
lines denote the cryogenic measurements performed at 10\,K. For the latter, 
the spectra are shifted by --0.1 for clarity. For montmorillonite, 
the room-temperature data have been measured without cryostat. 
For consistency, we display the 200\,K spectrum here. }
\label{overview}
\end{figure*}

\begin{table*}
\caption{Peak position $\lambda_p$, relative deviation between the positions at 
room temperature and 10~K $\Delta\lambda_p/\lambda_p$, and 10~K peak mass absorption coefficient $\kappa_p$ 
of the phyllosilicate FIR bands. The difference of $\kappa_p$ to the continuum 
baseline is additionally given in parentheses.}
\label{peaks} 
\begin{center}
\begin{tabular}{cccc|c|c|cccc|c|c}
%\hline
%\centering
 \multicolumn{6}{c|}{Talc} & \multicolumn{6}{c}{Picrolite} \\
 \multicolumn{4}{c|}{$\lambda_p$ ($\mu$m)} & $\Delta\lambda_p/\lambda_p$ (\%) & $\kappa_p$ (cm$^2$g$^{-1}$) & \multicolumn{4}{c|}{$\lambda_p$ ($\mu$m)} & $\Delta\lambda_p/\lambda_p$ (\%) & $\kappa_p$ (cm$^2$g$^{-1}$) \\
 300\,K & 200\,K & 100\,K & 10\,K & 300/10\,K & 10\,K & 300\,K & 200\,K & 100\,K & 10\,K & 300/10\,K & 10\,K \\ 
\hline
   --  &   --  & 26.5 & 26.5 & -- & 1000 (400) & 25.6 & 25.5 & 25.4 & 25.5 & 0.4 & 1700 (600) \\
 28.9 & 28.9 & 28.8 & 28.8 & 0.3 & 450 (340) & 26.8 & 26.8 & 26.8 & 26.7 & 0.4 & 1300 (800) \\
 32.0 & 32.0 & 31.9 & 31.9 & 0.3 & 110 (60) & 31.4 & 31.4 & 31.4 & 31.3 & 0.3 & 190 (140) \\
   --  & 34.2 & 34.0 & 34.0 & -- & 62 (30) & 32.7 & 32.7 & 32.6 & 32.6 & 0.3 & 2300 \\
 38.7 & 38.4 & 38.4 & 38.3 & 1.1 & 140 (110) & 42.9 & 42.6 & 42.5 & 42.4 & 1.2 & 530 \\
 43.2 & 43.2 & 43.0 & 42.9 & 0.7 & 51 (30) & 49.3 & 49.3 & 49.2 & 49.1 & 0.4 & 34 (25) \\
 56.5 & 56.5 & 56.0 & 56.0 & 0.9 & 130 (60) & 77.0 & 76.5 & 76.0 & 76.0 & 1.3 & 130 \\
 59.5 & 59.5 & 59.0 & 59.0 & 0.8 & 150 (90) & & & & & & \\
 98.5\footnotemark & 98.0 & 97.0 & 95.5 & 3.1 & 45 (20) & & & & & & \\
\hline
 \multicolumn{6}{c|}{Montmorillonite} & \multicolumn{6}{c}{Chamosite} \\
 \multicolumn{4}{c|}{$\lambda_p$ ($\mu$m)} & $\Delta\lambda_p/\lambda_p$ (\%) & $\kappa_p$ (cm$^2$g$^{-1}$) & \multicolumn{4}{c|}{$\lambda_p$ ($\mu$m)} & $\Delta\lambda_p/\lambda_p$ (\%) & $\kappa_p$ (cm$^2$g$^{-1}$) \\
 300\,K & 200\,K & 100\,K & 10\,K & 300/10\,K & 10\,K & 300\,K & 200\,K & 100\,K & 10\,K & 300/10\,K & 10\,K \\ 
\hline
 26.8 & 27.0 & 27.1 & 27.2 & -1.5 & 470 (100) & 29.4 & 29.4 & 29.2 & 29.1 & 1.0 & 1400 (550) \\
 29.0 & 29.0 & 29.0 & 29.0 & -- & 760 (440) & 35.2 & 35.2 & 35.0 & 34.6 & 1.7 & 940 (430) \\
 33.0 & 32.9 & 32.9 & 32.8 & 0.6 & 280 (45) & 41.0 & 40.5 & 40.0 & 40.0 & 2.5 & 440 (60) \\
 49.7 & 49.7 & 49.6 & 49.5 & 0.4 & 190 (55) &   --  & 42.5 & 42.5 & 42.5 & -- & 400 (80) \\
 60.0 & 60.0 & 60.0 & 60.0 & -- & 134 (8) & 49.5 & 49.5 & 49.0 & 49.0 & 1.0 & 310 (80) \\
   --  &  --  & 66.0 & 66.0 & -- & 129 (3) & 72.5 & 72.0 & 72.0 & 71.5 & 1.4 & 220 (90) \\
   --  &  72  &  72  &  72  & -- & 134 (15) & 77.0 & 76.5 & 76.5 & 76.5 & 0.6 & 190 (80) \\
  86  &  84  &  82  &  80  & 7.0 & 129 (15) & 91.5 & 90.5 & 89.0 & 89.0 & 2.8 & 220 (120) \\
 105 &  104 &  103 &  103 & 1.9 & 110 (25) &  --  &  --  & 108 & 108 & -- & 65 (10) \\
 & & & & & & 277 & 274 & 273 & 272 & 0.7 & 31 (19) \\
%\hline
\end{tabular}
\end{center}
\end{table*}

Figure \ref{overview} a--d displays the transmission spectra in the 25--110\,$\mu$m 
wavelength range, which were measured using the samples with the lower column 
densities (1.51\,mg/cm$^2$). For clarity, only the spectra measured at room temperature 
and at 10\,K are shown (the latter were shifted vertically).
The spectra in this range reveal significant differences between the spectroscopic 
properties of the four minerals. The picrolite spectrum, e.g., is characterized 
by comparably few but very strong and distinct FIR bands, while those of the other 
minerals are considerably weaker. The bands of the talc spectrum appear very sharp, 
while those of chamosite are much broader, and montmorillonite has a spectrum 
that is rather poor in structure beyond 30\,$\mu$m. The transmission plots allow 
a distinct delineation of the spectra with their bands of very different strengths, 
even without being able to constrain the continuum absorption, and they also illustrate
the experimental limitations (see below). Therefore we prefer the transmission scale here 
and give the peak mass absorption coefficients of the bands at 10~K and their 
difference to the estimated continuum absorption in tabular form (Table \ref{peaks}). 
The precision of the latter values is lower because of the uncertainty of the baseline. 

The noise level of the spectra is typically $\pm$1\% in transmission, 
but is higher in the range around 50\,$\mu$m, where the spectra measured with the two 
beamsplitters were merged and either of the two beamsplitters only provide low beam 
intensity. In general, the signal-to-noise ratio at this column density is too 
poor to reveal the weaker bands with sufficient clarity. Especially, the band expected 
for montmorillonite at $>$90\,$\mu$m is not detected in the room temperature spectrum 
and becomes only weakly visible at 10\,K. Instead, we see a continuously lowered 
transmission (to about T=0.85) at wavelengths $>$55\,$\mu$m. Additionally, most of the 
spectra show fringes in this wavelength range originating from multiple reflections at 
the pellet surfaces, which complicate the detection of weaker bands. 
For the continuum absorption behind the bands, i.e. the absolute transmission, 
the precision is even lower than the noise limit (e.g. caused by scattering and 
reflection compensation errors, see also Sect.\,\ref{mm}). Therefore, it cannot 
be determined precisely from these spectra, especially at the longer wavelengths, 
although some continuum absorption is detected for montmorillonite and chamosite. 

The temperature dependence of the band positions does on average not appear to be 
very strong, as has already been noted by Koike et al.\ (\cite{Koike82}). The 
expected broadening and the shift to longer wavelengths of the bands with 
increasing temperature (compare e.g.\ Henning \& Mutschke \cite{hemu97}, 
Koike et al.\ \cite{koimu06}, Suto et al.\ \cite{suto06}) is still quite 
clearly seen for some of the picrolite and chamosite bands, but is hardly 
detectable for most of the spectral features. Table \ref{peaks} presents 
the detailed peak wavelengths at all the temperatures measured. 
Peak wavelengths longer than 50\,$\mu$m are given with 0.5\,$\mu$m 
accuracy and values higher than 100\,$\mu$m, as well as those of heavily 
blended bands with 1\,$\mu$m accuracy. These data 
demonstrate that most peak wavelengths shift by less than 1~\% over the whole 
temperature range covered (the relative shifts are also given in the table). 
For some bands, no shift could be detected within the accuracy limits, which are 
between 0.1~$\mu$m and 1~$\mu$m depending on the strength of the band and the wavelength. 
Only for a few features is the shift considerably stronger. 
This is especially true for some bands in the 100~$\mu$m range. In some cases 
of weak and strongly shifting peaks, such as the montmorillonite 80~$\mu$m band 
and perhaps the chamosite 40~$\mu$m feature, the shift may be attributable 
to unresolved blending. 

We now discuss the spectra for each of the minerals separately, showing in 
addition the measurements at higher sample concentrations for ranges of weaker 
absorption. 

\subsection{Talc}

\begin{figure}
\flushleft
\includegraphics[width=9.5cm]{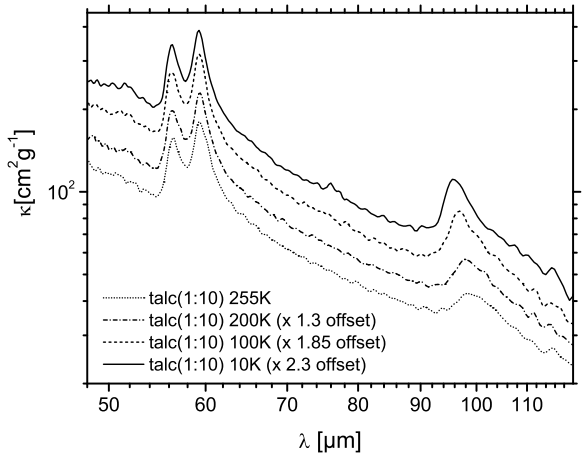}
\caption{Mass absorption coefficient in the FIR range 
for talc at different temperatures. The data have been measured 
at an enhanced powder concentration of 1:10. Note the change in 
width and band position of the $\sim$98\,$\mu$m band due to the 
cooling of the sample.}
\label{talc10}
\end{figure}

Comparing the spectra for talc displayed in Fig.\,\ref{overview} (a) with the 
studies by Zhang et al.\ (2006) and Hofmeister \& Bowey (2006), we note that 
(1) their results are consistent and (2) all of the bands 
reported at wavelengths beyond 25\,$\mu$m are present in our spectra as well, 
although the weak ones at 34\,$\mu$m and 43\,$\mu$m are hardly seen behind 
the noise especially in the room-temperature spectra. In addition, we see a 
weak band at about 95--100\,$\mu$m, which has not been noticed by the other 
authors although there are indications of its presence in their spectra
(Zhang, priv.\ comm.). 

For a closer look at this band and others at wavelengths beyond 50\,$\mu$m, 
we show in Fig.\,\ref{talc10} additional spectra measured at ten times 
higher talc column density. For an easier quantitative comparison, 
these spectra are displayed in terms of the mass absorption coefficient 
$\kappa = -ln(T) / \sigma$ calculated from the transmission T measured 
at temperatures between 255\,K and 10\,K. The newly discovered band is 
clearly resolved in these spectra. Its position is 98.5\,$\mu$m at 255\,K 
and 95.5\,$\mu$m at 10\,K with an FWHM decreasing from about 9\,$\mu$m to 4.5\,$\mu$m. 
In contrast to this rather strong temperature effect, the double feature 
at 56.5\,$\mu$m and 59.5\,$\mu$m changes much less with temperature. 
The positions of the latter features coincide well with the positions 
reported by Hofmeister \& Bowey (\cite{HB06}).

\footnotetext{This value corresponds to 255\,K (compare Fig.\,\ref{talc10}). }

%..............................................................................

\subsection{Picrolite}

The FIR data of minerals belonging to the serpentine group have been measured 
by both Hofmeister \& Bowey (\cite{HB06}) and Koike \& Shibai (\cite{KS90}). 
The former authors used lizardite and chrysotile, mentioning that their 
infrared spectra are similar. The latter measured a serpentine without 
further characterization. Picrolite (antigorite) is chemically identical to 
lizardite and chrysotile apart from the iron content, which for our 
sample is, however, very low and comparable to the minerals used by 
Hofmeister \& Bowey (\cite{HB06}). The crystal structure of picrolite and 
lizardite is in both cases monoclinic, as in most serpentines. Thus, we can 
expect that the IR spectrum displayed in Fig.\,\ref{overview} (b) will not 
show strong differences from the literature spectra. 

\begin{figure}
\flushleft
\includegraphics[width=9.5cm]{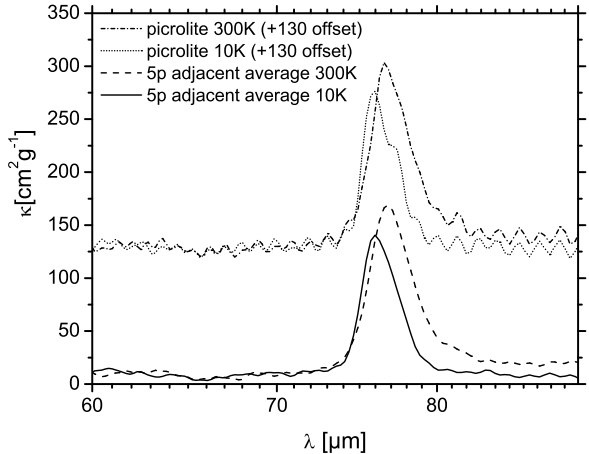}
\caption{Mass absorption coefficient spectra showing the $\sim$76$\mu$m 
picrolite band for temperatures of 300\,K and 10\,K.
The thin lines represent the original data -- affected by fringes --
while the bold lines denote the result of a 5-point-averaging procedure.}
\label{picsmoothed}
\end{figure}

Indeed we find band positions more or less identical to those of the lizardite 
measured by Hofmeister \& Bowey (\cite{HB06}) with one possible exception. 
The shoulder at 28.7\,$\mu$m (348 cm$^{-1}$) is not confirmed at 10\,K 
(at room temperature there is a weak uncertain structure), whereas other 
shoulders tend to be better-resolved at lower temperatures. 
See for instance the shoulder at about 27\,$\mu$m, which evolves into a 
distinct band at 10\,K. In the oscillator data of serpentine given 
by Koike \& Shibai (\cite{KS90}), all stronger bands are also present, 
most importantly the long-wavelength bands at 43, 48, and 79\,$\mu$m 
(43, 49, and 77\,$\mu$m in our data). Some shoulders are not represented 
by the oscillator data. However, the one at 28.7\,$\mu$m not seen in our 
spectra is again reported, and there are additional features at 37.5\,$\mu$m 
and 105\,$\mu$m not seen in any other reported spectra of this mineral class. 
Apart from these differences, we can confirm that the general features of 
the spectra are consistent for the minerals belonging to the serpentine group. 

Figure \ref{picsmoothed} displays the FIR part of the spectrum 
with the band at about 77-76\,$\mu$m. In the two lower spectra, the fringes 
were removed by simple smoothing (sliding average of 5 data points). 
This method works well for bands that are broad compared to the fringe 
spacing. For narrower bands we instead have either subtracted periodic 
functions, or we fitted the profile with Lorentzian profiles.

\subsection{Montmorillonite}
\label{mm}

Both the montmorillonite spectra published by Hofmeister \& Bowey (\cite{HB06}) 
and Koike \& Shibai (\cite{KS90}) have similarly shallow bands to ours. 
The band at 29\,$\mu$m is the most prominent longward of 25\,$\mu$m. 
The published data are in accordance with ours for features at 29\,$\mu$m, 
33\,$\mu$m (only Hofmeister \& Bowey \cite{HB06}), and about 50\,$\mu$m. 
However, our spectrum has an additional band at 26.8\,$\mu$m, which shifts to 
longer wavelengths at lower temperatures. On the other hand, we do not see 
the shoulders at 23.6\,$\mu$m and 39.4\,$\mu$m reported by the two previous 
papers at approximately identical positions. 

\begin{figure}
\flushleft
\includegraphics[width=9.5cm]{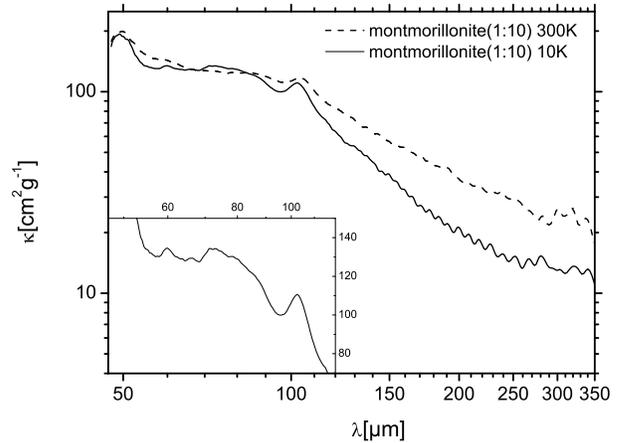}
\caption{Mass absorption coefficient spectra of the montmorillonite sample 
in the FIR range. The inset displays the 10\,K spectrum again on a linear scale 
to emphasize the individual absorption bands.}
\label{mont10}
\end{figure}

To look in more detail at the long-wavelength bands, we again measured 
spectra for a pellet with increased column density. Figure \ref{mont10} displays 
the mass absorption coefficient calculated from these spectra for 300\,K and 
10\,K. At the higher signal-to-noise ratio resulting from these measurements 
and especially by looking onto the 10\,K spectrum, it becomes clear that the 
montmorillonite spectrum is composed of a number of absorption bands that 
cover the whole wavelength range between 50 and 110\,$\mu$m. At room 
temperature, some of these bands merge into the very broad absorption 
structure peaking at about 86\,$\mu$m (consistent with both literature papers), 
whereas several additional peaks are seen at 10\,K. Especially a band at 
72\,$\mu$m strengthens continuously when cooling the sample, and the band at 
105\,$\mu$m separates from the shorter-wavelength ones, becoming more clearly 
visible and shifting to 103\,$\mu$m. Note that an indication of the latter 
band might also be present in Fig.\,5 of Hofmeister \& Bowey (\cite{HB06}). 

Because of these broad bands, the mass absorption coefficient at 120\,$\mu$m 
is, e.g., about four times higher than for talc (see also Fig.\,\ref{comp}), 
which allowed the decline of the absorption to be measured up to wavelengths of 
about 350\,$\mu$m. In contrast, this was not possible for talc and picrolite 
due to the small absorption signal. The uncertainty of the $\kappa$ values is 
about 5 cm$^2$g$^{-1}$ resulting from (1) influences of the cryostat setup, which 
caused slight differences between measurements with and without cryostat, 
which could not be clarified, (2) the uncertainty in smoothing of the fringes 
that may have been influenced by noise in the spectra, and (3) by the uncertain 
correction for pellet reflection losses. Embedding of 10\% of a silicate 
into a PE pellet can increase the reflectivity significantly, which is 
difficult to correct for quantitatively. 

For montmorillonite, the absorption even at these large wavelengths is still 
above the uncertainty level. Therefore, for room temperature we can confirm 
the finding of Koike \& Shibai (\cite{KS90}) of an approximate $\sim\lambda^{-1}$ 
behavior of the FIR continuum absorption. At 10\,K, however, the continuum 
decline is clearly seen to steepen and seems to approach an ordinary 
$\sim\lambda^{-2}$ slope (compare e.g.\ Mennella et al.\ 1998). 
The upturn of the 10\,K spectrum beyond 250\,$\mu$m is very uncertain, 
although Koike \& Shibai (\cite{KS90}) have reported such an upturn as well. 

The average absorption coefficient ($\alpha=\kappa\times\rho$, 
$\rho=2.5\pm0.2$ g\,cm$^{-3}$ being the density) that we derive 
for the 70--110\,$\mu$m spectral region is more than twice as high 
as the value given by Hofmeister \& Bowey (\cite{HB06}) 
(0.013\,$\mu$m$^{-1}$) but lower than the one given by Koike \& Shibai 
(\cite{KS90}) (3/4 Q$_{ext}$/a $\approx$ 0.045\,$\mu$m$^{-1}$). The 
latter difference is even greater at longer wavelengths. We tried to 
find out whether the inter-layer water present in montmorillonite may 
be responsible for these deviations and may generally have an influence 
on the FIR behavior. However, heating of the sample to 350$^\circ$C 
did not change the spectrum significantly. 
An alternative explanation for deviations between the experimental data 
could be provided by differences in grain size or shape, although simple calculations 
based on the optical constants by Koike \& Shibai (\cite{KS90}) would instead tend 
to predict absorption enhancement for both larger flake-like grains and also the 
compacted film used by Hofmeister and Bowey (2006) compared to the 1~$\mu$m sized 
grains measured by Koike \& Shibai (\cite{KS90}). 
This needs to be clarified by future investigations. 

%.............................................................................

\subsection{Chamosite}

\begin{figure}
\flushleft
\includegraphics[width=9.5cm]{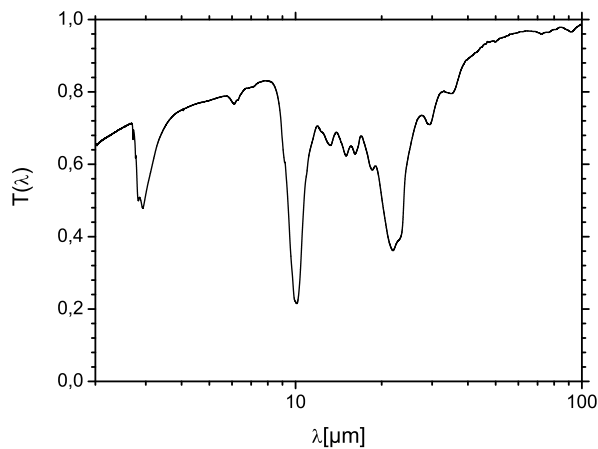}
\caption{Combined transmission spectrum of chamosite particles embedded in 
a KBr and a PE pellet at an identical column density of 0.3\,mg/cm$^2$. The
spectra were merged at 16.7\,$\mu$m.}
\label{cha_mir}
\end{figure}

The case for chamosite differs from those of the other minerals because
this mineral has not been measured yet at FIR wavelengths.
The mid-infrared spectra have been investigated by Zaikowski, Knacke, \& Porco (\cite{Zai75}),
Dorschner et al.\ (\cite{Dor78}), and Knacke \& Kr\"atschmer (\cite{Kna80}).
While Dorschner et al.\ (\cite{Dor78}) show only a 10\,$\mu$m profile,
the chamosite spectra displayed in the other two papers indicate a mixture 
of the samples used with a carbonate (identified by strong
bands at 7, 11, 14, and 30\,$\mu$m). We show for comparison the mid-infrared 
absorption spectrum of our chamosite powder embedded at a column density 
of 0.3\,mg/cm$^2$ into KBr and PE pellets (Fig.\,\ref{cha_mir}). The bands 
at wavelengths of 3 and 6\,$\mu$m are O--H stretching and bending vibrations. 
The FIR bands (at $>$50\,$\mu$m) appear very weak at this column density. The baseline slightly falling towards short wavelengths 
(near infrared) indicates scattering losses caused by chamosite grains
of $\sim$1\,$\mu$m size.

\begin{figure}
\flushleft
\includegraphics[width=9.5cm]{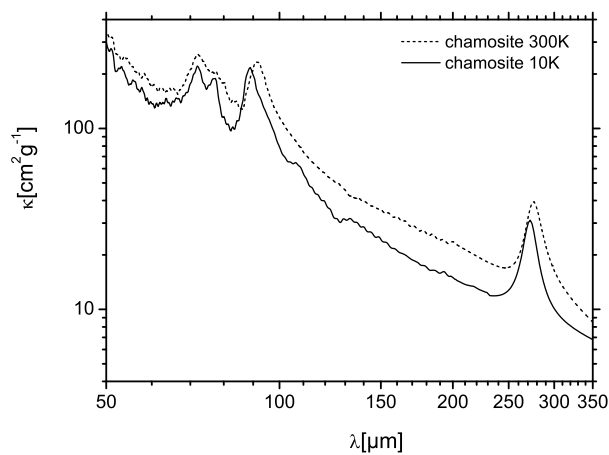}
\caption{Mass absorption coefficient spectra of the chamosite sample
in the FIR range. The spectra are a combinations of measurements at 
two different column densities (see text). The fringes in the 
original data have been removed by a sliding average method used 
up to a wavelength of about 200\,$\mu$m and by Lorentzian fitting 
for the 277\,$\mu$m band. The structure at about 130\,$\mu$m is due 
to imperfect compensation of a PE absorption band. }
\label{cha10}
\end{figure}

\begin{figure}
\flushleft
\includegraphics[width=9.5cm]{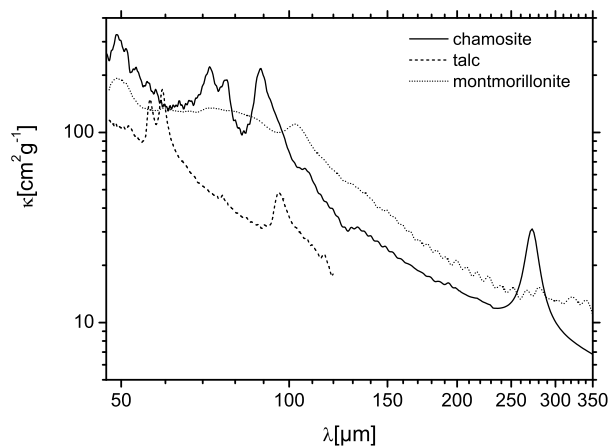}
\caption{Comparison of the mass absorption coefficients of three of the 
hydrous silicates measured at 10\,K. Picrolite is not shown because the 
pellet cooled to 10\,K had too low a column density to allow a derivation 
of the low (comparable to talc at room temperature) continuum absorption 
values. Picrolite has its longest wavelength band at 76\,$\mu$m at 10\,K, 
see Fig.\,\ref{picsmoothed}. }
\label{comp}
\end{figure}

We compared our FIR spectra (Fig.\,\ref{overview}d) with those obtained 
for chlorite by Koike et al.\ (\cite{Koike82}) and Koike \& Shibai 
(\cite{KS90}), because chamosite and other chlorites are isostructural and 
there are many similarities in their spectra. For the shorter-wavelength bands ($<$50\,$\mu$m), 
we find that the 29\,$\mu$m band of chlorite and the three minor bands between
40 and 50\,$\mu$m are also present for our chamosite, whereas chamosite's
additional stronger band at 35\,$\mu$m is not seen in the chlorite spectra.
The longer-wavelength bands are more clearly visible in Fig.\,\ref{cha10},
where the fringes have been smoothed out and the spectra have been
extended to 350\,$\mu$m using again a pellet of higher column density. 
At the signal-to-noise level of these data, we can still not confirm the 
chlorite bands at 53 and 58\,$\mu$m to have counterparts in the chamosite spectrum. 
On the other hand, chamosite has a strong band at 77\,$\mu$m that becomes 
clearly separated from the 72\,$\mu$m band at 10\,K and is not known from chlorite. 
The 91\,$\mu$m band does not strengthen at 10\,K as much as seen for chlorite by
Koike et al.\ (\cite{Koike82}), but shifts to 89\,$\mu$m. 
This may be influenced by a significant decrease in the continuum
absorption with lowering temperature, which is however not as strong as
for montmorillonite. Also the long-wavelength continuum absorption level is
lower in the case of chamosite (compare Fig.\,\ref{comp}), 
making the measured wavelength dependence uncertain. An exciting 
feature in the chamosite spectrum is the band at about 277\,$\mu$m, 
which certainly does correspond 
to the already mentioned 268\,$\mu$m feature of chlorite discovered 
by Koike \& Shibai (1990). These are the silicate lattice vibration 
with by far the lowest frequency known to us.

\begin{table*}
\caption{Peak position $\lambda_p$ and FWHM of the strongest FIR 
phyllosilicate bands at 10\,K.}
\label{FWHM}
\begin{center}
\begin{tabular}{ccc|ccc|ccc|ccc}
\hline
%\centering
\multicolumn{3}{c}{Talc} & \multicolumn{3}{c}{Picrolite} & \multicolumn{3}{c}{Montmorillonite} & \multicolumn{3}{c}{Chamosite} \\
$\lambda_p$ & FWHM & FWHM/$\lambda_p$ & $\lambda_p$ & FWHM & FWHM/$\lambda_p$ & $\lambda_p$ & FWHM & FWHM/$\lambda_p$ & $\lambda_p$ & FWHM & FWHM/$\lambda_p$ \\
($\mu$m) & ($\mu$m) & & ($\mu$m) & ($\mu$m) & & ($\mu$m) & ($\mu$m) & & ($\mu$m) & ($\mu$m) & \\
\hline
56.0 & 1.4 & 0.025 & 76.0 & 2.7 & 0.035 & 103 & 10.2 & 0.099 & 71.5 & 4.7 & 0.066 \\
59.0 & 1.8 & 0.031 &      &     &       &     &      &       & 76.5 & 3.7 & 0.048 \\
95.5 & 4.5 & 0.047 &      &     &       &     &      &       & 89.0 & 6.6 & 0.074 \\
     &     &       &      &     &       &     &      &       &  272 &  17 & 0.062 \\
\hline
\end{tabular}
\end{center}
\end{table*} 

\begin{figure*}
\centering
\includegraphics[width=12cm]{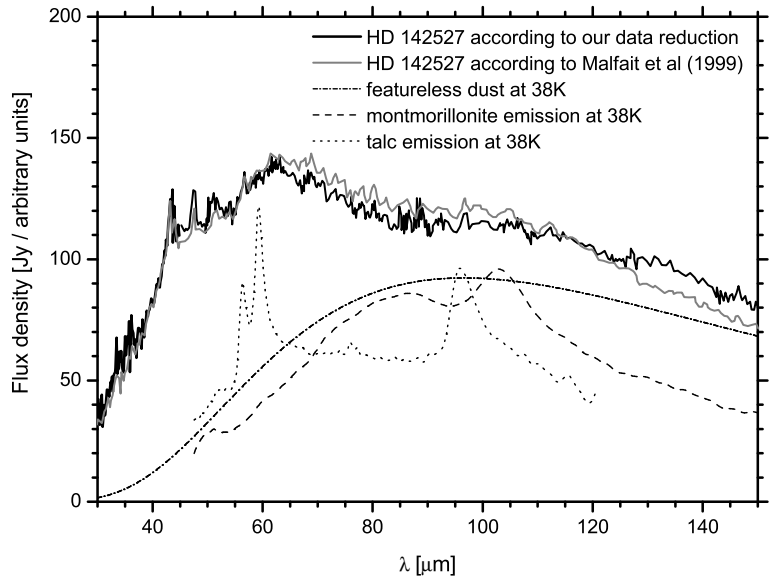}
\caption{The ISO spectrum of HD 142527 (solid lines) compared to the 
emission spectra of cold montmorillonite (dashed line) and talc dust 
(dotted line). The latter have been calculated by multiplying the mass 
absorption coefficients $\kappa$ of montmorillonite and talc {at 10\,K } 
with a blackbody function for the best-fitting temperature, T=38\,K. 
Featureless dust ($\kappa_{FIR}$ $\propto$ 1/$\lambda$; dash-dotted line) 
is an alternative carrier of the broad emission in the 100\,$\mu$m region.}
\label{ast}
\end{figure*}

%-----------------------------------------------------------------------------

\section{Comparison with astronomical spectra \label{sec:astro}}

\subsection{A broad montmorillonite band around 100\,$\mu$m
in the young stars HD 142527 and HD 100546?}

Malfait et al.\ (1999) present quite convincing evidence of 
montmorillonite grains in the dust disk of the Herbig Ae/Be star 
HD 142527. They modeled the ISO SWS and LWS spectra of this 
object by means of a radiative transfer calculation, assuming 
the following dust components: amorphous and crystalline silicates, 
H$_2$O-ice, montmorillonite, and a small fraction of FeO. 
The dust temperature in the disk was found to be in the range 
30 -- 60\,K. 

In Fig.\,\ref{ast} we present an independent comparison between the ISO
spectrum of HD 142527 and montmorillonite dust. A temperature of 38\,K 
was found to allow the best fit to the observed spectrum, although the 
band profile in the 100\,$\mu$m range depends much more strongly on 
the material's optical properties than on the chosen temperature. 
We would like to point out the absence of a narrow 105\,$\mu$m band 
in the ISO spectrum. 

The ISO spectrum was reduced with the ISO spectral analysis package ISAP, 
version OLP 10.1. The differences between 
the two versions of the ISO spectra are negligible in the MIR range.
In the 60-160\,$\mu$m region, the spectrum published by Malfait et al.\
has a steeper overall gradient. It can be seen from Fig.\,\ref{ast}
that the presence or absence of a band in the 100$\mu$m region does
not depend on the version of the spectrum to which our laboratory data
are compared. However, the details of the profile of the excess emission 
around 100\,$\mu$m do depend on the preconditioned data reduction procedure.
Observations of  HD 142527 with an entirely different instrument (e.g.\
{\em Herschel}\/-PACS) will help to clarify the real shape of the 
spectral energy distribution of this object. 

The montmorillonite spectrum clearly differs from the one shown
by Malfait et al.\  This is mainly due to the narrow 103\,$\mu$m band
of montmorillonite, emerging as an increasingly sharp structure at low 
temperatures as shown in Fig.\,4. At 10\,K, the FWHM of this band is 
10.2\,$\mu$m. We conclude from this comparison that cold montmorillonite 
dust cannot be the carrier of the broad $\sim$100\,$\mu$m shoulder 
in the spectrum of HD 142527. 
Figures \ref{comp} and \ref{ast} demonstrate that the same
conclusion holds for talc and chamosite as well. Both of these hydrous 
silicates have at least similarly strong and sharp bands in the 
$\sim$100\,$\mu$m wavelength range 
(at 89 and 95.5\,$\mu$m with FWHM values of 6.6 and 4.5\,$\mu$m, respectively, 
see also Table \ref{FWHM}). 

The finding that all of the three-layer phyllosilicates investigated 
here (picrolite is of the two-layer type, see Sect.\,2.1) have sharp bands in 
the $\sim$100\,$\mu$m wavelength range makes us confident that the 
montmorillonite 103\,$\mu$m feature is not a peculiarity of our 
montmorillonite sample, but indeed characteristic of montmorillonite. 
As noted before, an indication of the band is present in the room temperature 
spectrum presented by Hofmeister \& Bowey (\cite{HB06}), and may even be present 
in the 2\,K spectrum of Koike et al.\ (\cite{Koike82}). In addition, this 
band is also seen in spectra of the already mentioned ``bentonite''. 

Only a material that has a much smoother FIR spectrum than the phyllosilicate 
spectra presented here, even at low temperatures, can account for the
observed FIR spectral energy distributions of HD 142527. 
This fact has already prompted Verhoeff et al.\ (2005) to fit the FIR part 
of the spectrum with iron sulfide (FeS). In contrast to montmorillonite,
iron sulfide is essentially featureless in the wavelength region beyond 
60\,$\mu$m; it is characterized by an almost constant real part and 
a monotonically rising imaginary part of the complex index of refraction 
in this wavelength range, according to Henning \& Mutschke (\cite{hemu97}). 
However, we want to stress that {\em any}\/ kind of dust 
with a 1/$\lambda$ dependence of its absorption coefficient in the FIR 
could provide a good match of the $\sim$100\,$\mu$m shoulder 
(see Fig.\,\ref{ast}) if it were characterized by a narrow range of 
dust temperatures (about 35--40\,K), corresponding to a region in the outer 
disk where this temperature is prevailing. 
Alternatively, dust with a `steeper' absorption coefficient in the FIR
(e.g.\ $\propto$ 1/$\lambda^2$), in combination with a wider range
of dust temperatures (hence a broader spectral energy distribution
of underlying blackbodies), may account for the FIR spectrum of 
HD 142527. 

The same arguments apply to the dust around the Herbig Ae/Be star
HD 100546. Bouwman et al.\ (2003) point out that there is a broad
shoulder, extending from 85--125\,$\mu$m, in the ISO-LWS spectrum of this
object, although they note that a change in the supposed
underlying continuum `may change both the strength and width of the
feature or may even cause it to disappear'. In Fig.\,7 of their paper, 
they compare this emission feature to the structure in HD 142527 
discussed above and find widths and positions to be similar for both bands, 
hence too wide to be ascribed to the phyllosilicates studied here.

\subsection{Other objects with broad bands around 100\,$\mu$m}

Chiavassa et al.\ (2005) studied a sample of 32 low and intermediate 
mass protostars (mostly class\,0 and Herbig Ae/Be objects). 
Out of this sample, the spectra of 17 sources show a broad band feature 
between 90 and 110\,$\mu$m. The dust temperature in these objects is 
below 50\,K, except for emission from the innermost regions. 
Calcite is suggested as a possible carrier of this feature, which seems 
to be supported by the laboratory measurements of Kemper et al.\ (2002), 
which show a peak at 93\,$\mu$m with a width (FWHM) of 16\,$\mu$m. 

A part of the above-mentioned sources, mainly the class\,0 protostars, show 
a broad band peaking beyond 100\,$\mu$m. According to Posch et al.\
(2007), calcite cannot account for that emission band since its 
lowest frequency band peaks at too short a wavelength. 
Chiavassa et al.\ ponder the possibility that hydrous silicates
could account for the $>$100\,$\mu$m bands/shoulders of
their class\,0 protostars, but conclude -- on the basis of the 
optical constants published by Koike \& Shibai (1990) --
that the bands of these hydrous silicates are much too broad. 
On the contrary, our data -- revealing sharp phyllosilicate 
features in the FIR -- lead us to the opposite conclusion: 
the width of the phyllosilicate emission bands so far known 
to result for dust temperatures of 10--100\,K is definitely 
{\em smaller}\/ than required by the observations. 
If any of the phyllosilicates discussed in the present paper
were present in significant amounts in protostars, rather
narrow FIR bands would reveal their presence. 

%-----------------------------------------------------------------------------

\begin{acknowledgements}
We thank Gabriele Born and Walter Teuschel, Jena, for the
sample preparation and for help with the cryogenic measurements.
HM acknowledges support by DFG grant Mu \mbox{1164/6.} 
TP, FK, and AB acknowledge support by the Austrian `Fonds zur F\"orderung 
der wissenschaftlichen Forschung' (FWF; project number 
P18939-N16).
\end{acknowledgements}

%======================================================


\begin{thebibliography}{}

\bibitem[2002]{BA02}
Bowey, J.E. \& Adamson, A.J., 2002,
\mnras, 334, 94

\bibitem[2003]{bouwman03}
Bouwman, J., de Koter, A., Dominik, C., Waters, L.B.F.M., 2003,
A\&A, 401, 577

\bibitem[2005]{Chia05}
Chiavassa, A., Ceccarelli, C., Tielens, A.G.G.M., Caux, E., \& Maret, S., 
2005, A\&A, 432, 547

\bibitem[1978]{Dor78}
Dorschner, J., Friedemann, C., \& G\"urtler, J., 1978, AN, 299, 269

\bibitem[1995]{hemu97} 
Henning, Th. \& Mutschke, H., 1997, \aap, 327, 743

\bibitem[2006]{HB06}
Hofmeister, A.M. \& Bowey, J.E., 2006, MNRAS, 367, 577

\bibitem[1998]{jae98}
J\"ager, C., Molster, F.J., Dorschner, J., Henning, Th., Mutschke, H., 
\& Waters, L.B.F.M., 1998, \aap,  339, 904

\bibitem[2002]{Kemp02}
Kemper, F., Molster, F.J., J\"ager, C., Waters, L.B.F.M., 
2002a, A\&A 394, 679

\bibitem[1980]{Kna80}
Knacke, R.F. \& Kraetschmer, W., 1980, \aap, 92, 281

\bibitem[1982]{Koike82}
Koike, Ch., Hasegawa, H., \& Hattori, T., 1982, ApSS, 88, 89

\bibitem[1990]{KS90}
Koike, Ch. \& Shibai, H., 1990, \mnras, 246, 332

\bibitem[2006]{koimu06}
Koike, Ch., Mutschke, H., Suto, H., Naoi, T., Chihara, H., et al., 2006,
\aap, 449, 683

\bibitem[1964]{Kroenert64}
Kr\"onert, W., Schwiete, H.E., \& Suckow, A., 1964,
Naturwiss., 51, 4, 85

\bibitem[2007]{Lisse07}
Lisse, C.M., Kraemer, K.E., Nuth, J.A., Li, A., \& Joswiak, D., 2007, 
Icarus, 191, 223

\bibitem[1999]{M99}
Malfait, K., Waelkens, C., Bouwman, J., de Koter, A., Waters, L.B.F.M., 1999,
A\&A, 345, 181

\bibitem[1998]{Mennella98}
Mennella, V., Brucato, J.R., Colangeli, L., Palumbo, P., Rotundi, A., 
Bussoletti, E., 1998, \apj, 496, 1058

\bibitem[1985]{MK85}
Mooney, T. \& Knacke, R.F., 1985, Icarus, 64, 493

\bibitem[1975]{Nagy75}
Nagy, B., 1975, Carbonaceous Meteorites, Elsevier, Amsterdam

\bibitem[2005]{PKM05}
Posch, Th., Kerschbaum, F., Richter, H., Mutschke, H., 2005, 
ESA-SP-577, 257

\bibitem[2007]{PM07}
Posch, Th., Baier, A., Mutschke, H., Henning, Th., 2007, 
\apj, 668, 993

\bibitem[1985]{SW85}
Sandford, S.A. \& Walker, R.M., 1985,
\apj, 291, 838

\bibitem[2005]{Smith05}
Smith, J.B., Dai, Z.R., Weber, P.K., et al., 2005, 
LPI conf. abstr., 36, 1003

\bibitem[2006]{suto06}
Suto, H., Sogawa, H., Tachibana, S., Koike, C., Karoji, H., et al., 2006,
\mnras, 370, 1599

\bibitem[1977]{T77}
Toon, O.B., Pollack, J.B., \& Sagan, C., 1977,
Icarus, 30, 663

\bibitem[2005]{V05}
Verhoeff, A.P., Min, M., de Koter, A., et al., 2005,
Protostars and Planets V, 8465

\bibitem[1975]{Zai75}
Zaikowski, A., Knacke, R.F., \& Porco, C.C., 1975, 
\apss, 35, 97

\bibitem[2006]{Zhang06}
Zhang, M., Hui, Q., Lou, X.-J., et al., 2006, 
Amer.\ Min.\, 91, 816

\end{thebibliography}
\end{document}